\documentclass{appolb}
\usepackage{amsmath}
\usepackage{graphicx}
\usepackage{subcaption}
\usepackage{cite}
\usepackage{wrapfig}
\usepackage{xcolor}
\begin{document}
\title{Pairing dynamics in low energy nuclear collisions%
\thanks{Presented at the XXXVI Mazurian Lakes Conference on Physics, Piaski, Poland, September
1-7, 2019.}
}

\author{M.~C.~Barton$^{a}$, S.~Jin$^{b}$, P.~Magierski$^{a,b}$, K.~Sekizawa$^{c}$, G.~Wlaz{\l}owski$^{a,b}$ and A.~Bulgac$^{b}$
\address{$^a$ Warsaw University of Technology, Poland\\ $^b$ University of Washington, USA\\ $^c$ Niigata University, Japan}} 

\maketitle
\begin{abstract}
Superfluidity is a generic feature of various quantum systems at low temperatures and 
it is in particular important for the description of dynamics of low energy nuclear reactions. 
The time-dependent density functional theory (TDDFT) %\cite{PhysRevLett.119.042501,2018AcPPB..49..281M,PhysRevLett.116.122504} 
is, to date, the only microscopic method which takes into account in a consistent way
far from equilibrium dynamics of pairing field and single-particle degrees of freedom.
The local version of TDDFT, so called TDSLDA, is particularly useful for 
the description of nuclear reactions and is well suited for
leadership class computers of hybrid (CPU+GPU) architecture.  
The preliminary results obtained for collisions involving both medium-mass and heavy nuclei
at the energies around the Coulomb barrier are presented. 
%The symmetric collisions are used to study how solitonic excitations effect the fusion threshold \cite{PhysRevLett.119.042501}, while %the asymmetric collisions are used to study quasi fission. 
\end{abstract}
  
\section{Introduction}
The Time-Dependent Superfluid Local Density Approximation (TDSLDA) is a versatile tool to investigate a variety of phenomena 
involving superfluidity in Fermi systems including atomic nuclei. 
TDSLDA originates from time dependent density functional theory, which become nowadays a standard theoretical tool 
for studies of interacting many-body Fermi systems and offers a universal approach to the
quantum many-body dynamics (see \cite{Ullrich:2012,Marques:2012,Oni2002} and references therein).
The superfluid extension of TDDFT has been triggered by the discovery of high-temperature superconductivity
and resulted in the creation of nonlocal TDDFT for superconductors \cite{Oli1988, Wack1994}. It has been possible however 
to formulate the problem using local pairing field \cite{Kur1999}. The justification for the so-called   
SLDA (Superfluid Local Density Approximation) has been developed in a series of papers 
(see, e.g., a review \cite{bulgac2013} and references therein)
and it has been shown to be very accurate for nuclei and cold atomic gases\cite{2016arXiv160602225M,Bulgac:2019sac}. 
The recent studies of various low-energy 
nuclear reactions and in particular induced fission 
\cite{stetcu2011,stetcu2015,bulgac2016,PhysRevLett.119.042501,Grineviciute:2017jea,Bulgac2019} has proved 
that the TDSLDA is capable of describing nuclear processes, where pairing correlations play a crucial role.

In the following we will present selected aspects of superfluid dynamics in low-energy nuclear reactions. 
We consider two particular cases associated with two different types of collisions.
In the first one we will consider a mass-symmetric collision of superfluid nuclei which have different phases of the pairing field, giving rise to solitonic excitation between two colliding nuclei. The second example concerns collisions of a light and a heavy superfluid nuclei and 
is oriented towards investigation of an effect of superfluidity on the quasi-fission process in reactions reading to the formation of superheavy elements (SHEs).
%by comparing the properties of the final fission fragments from superheavy nuclear synthesis. 
%The quasi fission mechanism is one of the main obstacles encountered during the creation of SHE and therefore understanding of
%its role and dependence on pairing dynamics is od paramount importance.
%process by which a nucleus formed from a collision, fissions before it can form a compound nucleus.\\
% \indent The paper is set out as follows. Section 2 will give a brief description of how the ground states are obtained within 
% TDSLDA. Section 3 will present current preliminary results obtained for applications of TDSLDA to investigate solitonic excitations and quasi fission. Section 4 will give a brief summary of the main results, discuss possible extension of this work in the future. 

\section{Computation Background}
	There are a variety of computational methods to approximate the static solution of TDSLDA. They usually involve 
	a number of diagonalizations of the Hamiltonian matrix, which is a quite computationally demanding task as the size
	of the matrix is of the order of the lattice size.
	The one employed in this work is known as the Conjugate-Orthogonal Conjugate-Gradient (COCG) method (see Ref. \cite{PhysRevC.95.044302} for a detailed description). Its main advantage is that it does not require diagonalizations
	during the iteration process, namely, both normal and anomalous densities are constructed through evaluation of the 
	Green's function of the problem. 
	
	The typical procedure applied in the context of the presented studies is the following:
\begin{enumerate}
 \item Two nuclei are placed inside a box, symmetrically at the relative distance of $40$\,fm, and an external potential $V_{\rm ext}(\boldsymbol{r})\simeq V_0|x|$ which generates constant force pointing towards center of the box $x=0$
is used to counteract the Coulomb repulsion. The grid spacing is $1.25$\,fm in all directions, and the box size 
is $80 \times 25 \times 25$\,fm$^{3}$.
 \item Self-consistent iterations are executed using COCG code~\cite{PhysRevC.95.044302}, and the density solution found is inputted into a diagonalisation code, in order to extract the quasi-particle wave functions.
%Although one could in theory solve the problem directly using the diagonalisation code, without the need for the COCG code, the COCG %methods allows one to get a good approximation for the solution while being more computationally efficient. 
 \item The wave functions are then evolved solving TDSLDA equations, which are formally equivalent 
to the time-dependent Hartree-Fock-Bogoliubov equations (see Ref.~\cite{2016arXiv160602225M} for a review). 
\end{enumerate}	
For a thorough description of this process the reader is directed towards the supplementary material of Ref. \cite{PhysRevLett.119.042501}.
In this work the full Skyrme functional (SLy4) is used, including the spin-orbit term. 
%The accuracy of the evolution may be estimated by considering the total energy conservation
%as a function of time. It has been found that deviation from the initial value is less than
%$1$\,MeV (depending on the energy cutoff applied) for trajectories of length of the order of $10^4$\,fm/$c$ %\cite{Grineviciute:2017jea}.

%It can be generated by a zero range phenomenological interaction with 9 parameters. The parameterisation used in this work is Sly4. 
%The extension to the SeaLL1 functional will be also , which is dependent on only 7 parameters (see Ref. \cite{PhysRevC.97.044313} %for derivation of this functional). This should allow us to determine whether the analysis presented in this paper is independent of %the functional being used.

\section{Nuclear Collisions within TDSLDA}
Nuclear collisions simulated within the TDSLDA framework take into account dynamics of
both single particle and pairing degrees of freedom during the collision process \cite{magierski2018}.
The pairing field is a complex field and therefore its excitations consist both of variations of the magnitude and the phase, i.e.~$\Delta ( { \bf r }, t) = \lvert \Delta ( { \bf r } , t )   \rvert e^{ i \phi( { \bf r } , t ) } $. In particular 
the combination of both phase
and magnitude variations may lead to a long-lived solitonic excitations observed in superfluid systems \cite{wlazlowski2018} and also
predicted recently in the case of nuclear collisions \cite{PhysRevLett.119.042501}.

This section will now provide a mainly qualitative summary of the results of our investigations to date. 
It is split into two subsections. In one subsection we discuss the results obtained for solitonic excitations, and in the other one we discuss pairing
dynamics in the context of quasi-fission.

\subsection{Solitonic Excitations}
Collisions of two superfluid nuclei may differ not only by the magnitude of the pairing gap but also by the phase.
The latter quantity is not controlled in nuclear systems but may lead to observable effects. Namely, a nonzero phase difference
creates a long-lived solitonic structure between colliding nuclei, where part of the kinetic energy is stored.
This in turn creates an additional barrier for fusion which a projectile needs to overcome.
In order to investigate possible consequences of this effect medium mass-symmetric collisions are considered.
In this process the magnitude of pairing field of both nuclei is the same and the only effect comes from the pairing
phase difference. Therefore 
the mass-symmetric head-on collisions of $^{90}$Zr+$^{90}$Zr and $^{96}$Zr+$^{96}$Zr were investigated within the TDSLDA framework,
with the goal of determining the change of the barrier height for capture as a function of the relative phase difference
$\Delta\phi$ between colliding nuclei.
%The phase difference of the pairing field generates the solitonic excitation during the collision giving rise to an increase of the %Coulomb barrier. 
By comparing $^{90}$Zr+$^{90}$Zr (zero pairing gap) to $^{96}$Zr+$^{96}$Zr ($ \approx $ 1 MeV pairing gap) one may deduce the magnitude of this increase relative to the magnitude of the pairing gap.
The previously reported results in Ref.~\cite{PhysRevLett.119.042501} were performed with the Fayans functional without the spin-orbit term. In the present work the full SLy4 functional has been applied.
The timescale under which it was checked whether nuclei split was approximately $10^4$\,fm/$c$. 
It is observed that $^{96}$Zr+$^{96}$Zr collisions indeed produce the gauge-angle dependent 
barrier for capture, confirming the earlier results. Consequently the effective barrier for capture in $^{96}$Zr+$^{96}$Zr is enhanced as compared to $^{90}$Zr+$^{90}$Zr, where no effect is observed. The SLy4 functional produces weaker enhancement of the barrier compared to the value reported in \cite{PhysRevLett.119.042501}, in agreement with empirical analysis of~\cite{PhysRevC.97.044611}.
The details of this study will be reported elsewhere.
\begin{figure}[htb]
  \vspace{-11pt}
\centerline{
\begin{subfigure}{0.5\textwidth}
  \centering
  \includegraphics[width=\linewidth]{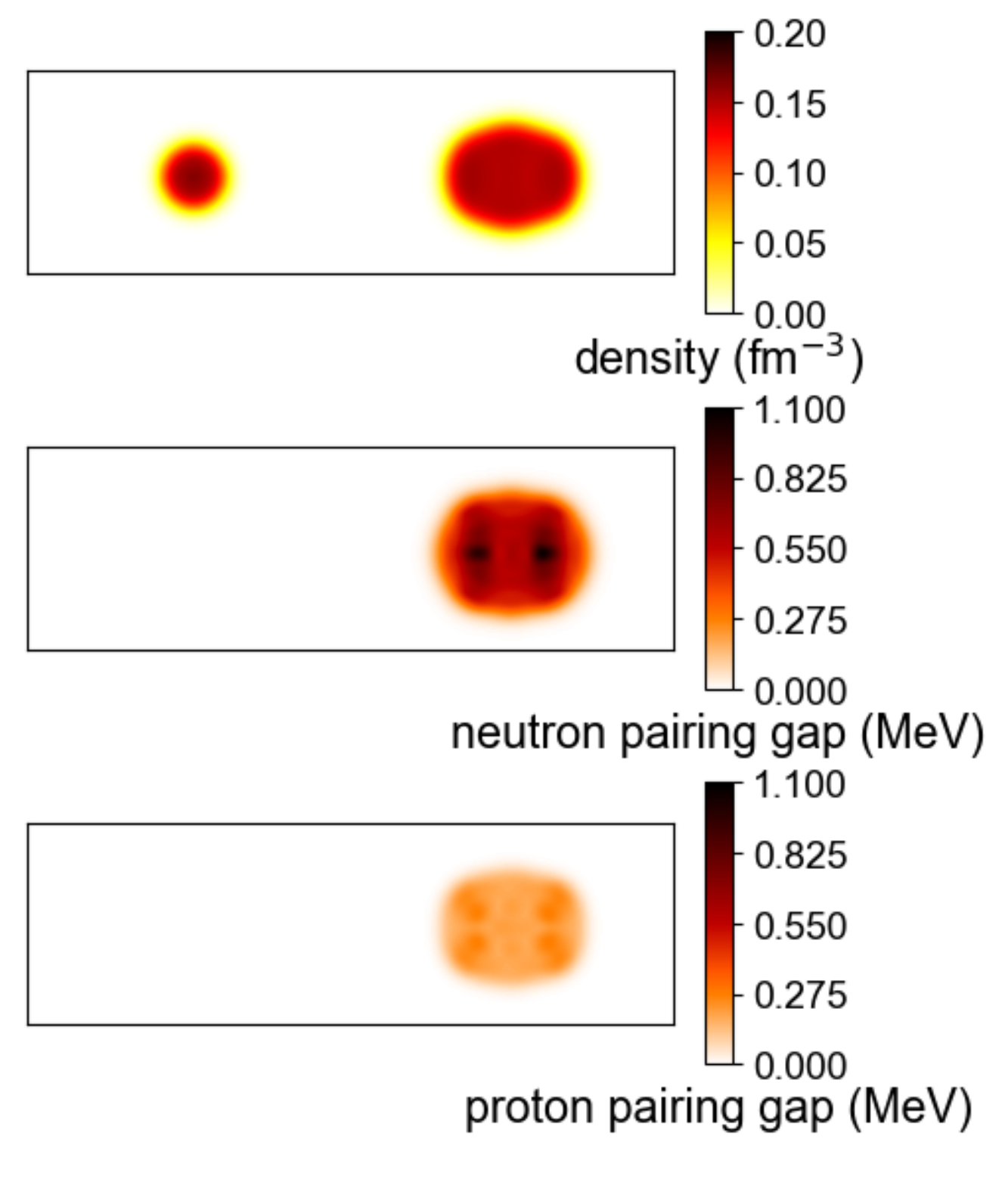}
 % \caption{t=0 fm / c}
  \label{fig:sfig1}
\end{subfigure}%
\begin{subfigure}{0.5\textwidth}
  \centering
  \includegraphics[width=\linewidth]{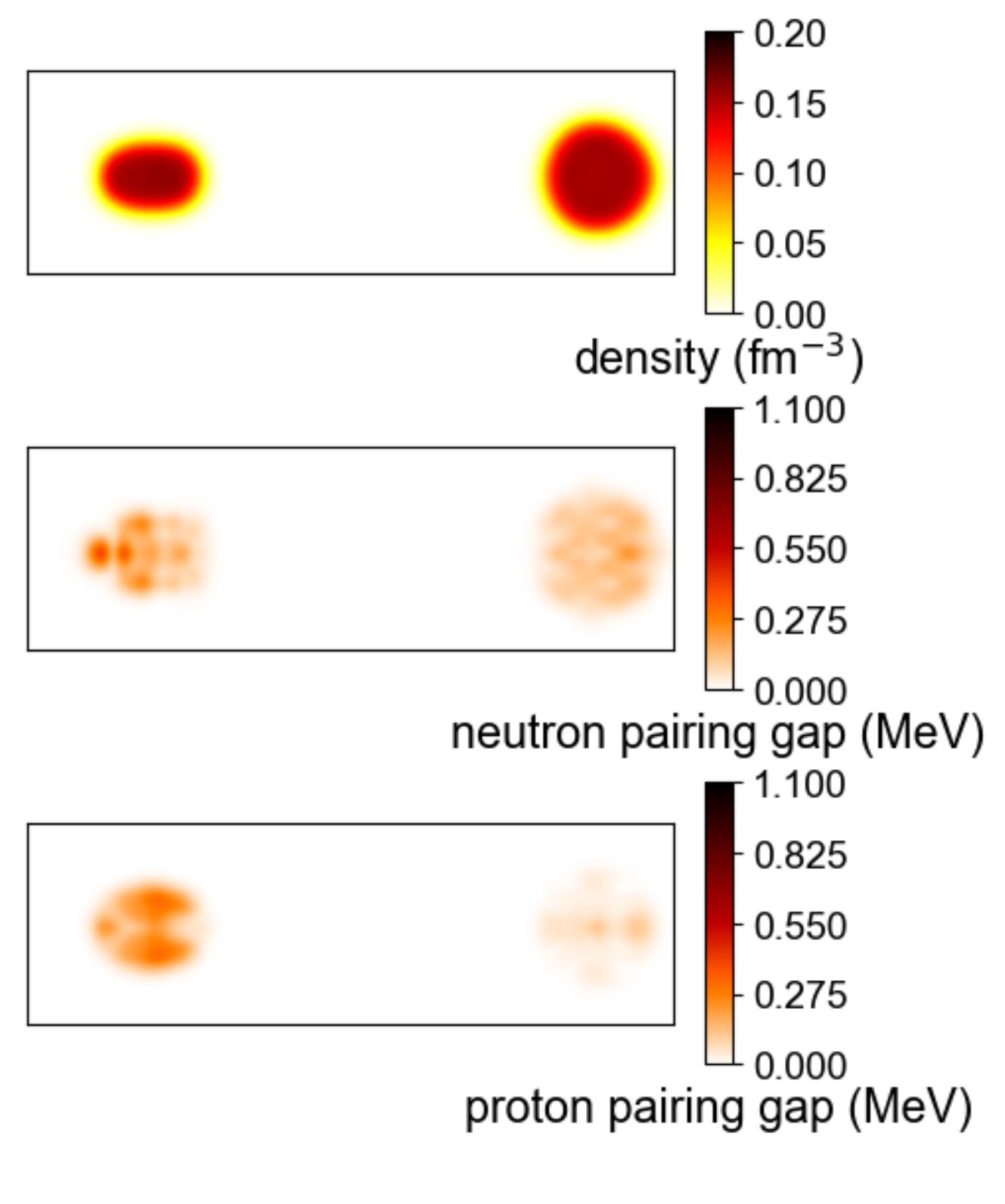}
%  \caption{t $\approx 9600 $ fm / c }
  \label{fig:sfig3}
\end{subfigure}
}
  \vspace{-11pt}
\caption{Snapshots before ($t = 0$~fm/$c$, left three) and after ($t \approx 10200$\,fm/$c$, right three) of a tip collision of $^{48}$Ca+$^{252}$Cf. The centre of mass energy in this example is 230\,MeV.}
\label{Fig:F2H}
\end{figure}
\subsection{Mass-asymmetric collisions}
The mass-asymmetric collisions of $^{48}$Ca+$^{252}$Cf and $^{50}$Ti+$^{252}$Cf offer the possibility to investigate the influence of pairing on the mechanism of the formation of SHEs.
Here we report preliminary results concerning tip collisions within the CM energy range 200--300\,MeV for $^{48}$Ca+$^{252}$Cf, and 220--300\,MeV for $^{50}$Ti+$^{252}$Cf.
Notice that the projectile $^{48}$Ca is essentially in a normal state, while $^{252}$Cf possess nonzero
pairing gap for both protons and neutrons. The projectile $^{50}$Ti possess a nonzero proton pairing. Therefore these two reactions represent interesting cases to investigate the collision of normal and superfluid systems
and to compare it to the case of superfluid on superfluid collisions.

From the results we found that the heavy fragment to be close to a doubly-magic nucleus,
regardless of the nuclei involved in the collision and energy. Although the split was similar in terms of proton and neutron numbers, it was generally found that the contact time of the two nuclei was several $1000$\,fm/$c$ longer for
$^{48}$Ca+$^{252}$Cf. This occurs when the amount of kinetic energy that the heavy fragment carries away is noticeably smaller (5--10\,MeV).
This is similarly seen in TDHF calculations. Visually comparing the collisions of $^{48}$Ca+$^{252}$Cf (Fig.~1) and $^{50}$Ti+$^{252}$Cf (not shown), it appears that the resultant nucleus is more elongated for $^{48}$Ca+$^{252}$Cf before fissioning. The larger elongation should reduce the magnitude of the Coulomb repulsion between the fragments, thereby reducing their kinetic energies after fission.

Comparing the contact times for the two cases to similar collisions in TDHF calculations, it is found that the contact time is approximately 2--3 times longer for TDSLDA.
The important effect of the $^{48}$Ca+$^{252}$Cf collision consists of the pairing transfer from the superfluid 
heavy nucleus to the initially-normal light projectile. Such an effect is surprising at first sight as it corresponds to inducing 
pairing correlations in the system which is heated up due to collision and is otherwise in the normal state. However, one needs to take into account
that a nucleus is in a nonequilibrium state. Also, the pairing properties in the nuclear system depends on
the level density at the Fermi surface, which may be changed at finite excitation energy and thus allow
for pairing correlations to set in. A similar effect known as ``pairing reentrance" has been predicted in hot rotating 
atomic nuclei \cite{kammuri,BFV}.
\vspace{-13pt}
\section{Conclusion}
%We now conclude with a summary of the main results. 
$^{48}$Ca+$^{252}$Cf and $^{50}$Ti+$^{252}$Cf collisions were performed within the TDSLDA framework, and reaction dynamics were compared to investigate the quasi-fission process. It was found that contact times for $^{48}$Ca+$^{252}$Cf were generally significantly longer than that obtained for $^{50}$Ti+$^{252}$Cf. This coincided with a reduction in the kinetic energy after collision for $^{48}$Ca+$^{252}$Cf. 
It was also observed that the superfludity is transferred to the initially-normal projectile $^{48}$Ca as a result of non-equilibrium single-particle and pairing field dynamics.
In the future we plan to reperform the same calculations but with a different orientations of the $^{252}$Cf nucleus. \\
\indent Also $^{90}$Zr+$^{90}$Zr and $^{96}$Zr+$^{96}$Zr collisions were compared to see how a solitonic excitation created between the two fragments would affect the fusion threshold energy. By creating a $ \pi $ phase difference between the two superfluid nuclei it was found that this threshold increased by $^{96}$Zr+$^{96}$Zr by 5\,MeV compared to no phase difference, but no increase was found for $^{90}$Zr+$^{90}$Zr. In the future we plan to reperform the calculations with a larger pairing gap. Finally we plan to 
%The extension to the SeaLL1 functional will be also , which is dependent on only 7 parameters (see Ref. \cite{PhysRevC.97.044313} for derivation of this functional). This should allow us to 
determine whether the analysis presented in this paper is independent of the functional being used.
\vspace{-13pt}
\section*{Acknowledgements}
    This work was supported by the Polish National Science Centre (NCN) under Contracts No. UMO-2016/23/B/ST2/01789 and
    UMO-2017/27/B/ ST2/02792. We acknowledge PRACE for awarding us access to resource Piz Daint based in Switzerland at Swiss National Supercomputing Centre (CSCS), decision No. 2017174125 and No. 2018194657. We also acknowledge the Global Scientific Information and Computing Center, Tokyo Institute of Technology, for resources at TSUBAME3.0 (Project ID: hp190063).
    
This research used resources of the Oak Ridge
Leadership Computing Facility, which is a U.S. DOE Office of Science
User Facility supported under Contract No. DE- AC05-00OR22725 and of
the National Energy Research Scientific computing Center, which is
supported by the Office of Science of the U.S. Department of Energy
under Contract No. DE-AC02-05CH11231. 

\bibliographystyle{unsrt}
\bibliography{paper2}

\end{document}